\documentclass[aps,prl,twocolumn]{revtex4}
\usepackage{amsmath, amsthm, amssymb,graphicx}
\begin{document}
\title{Thin Spectrum States in Superconductors}
\author{Jasper van Wezel and Jeroen van den Brink}

\affiliation{
Institute-Lorentz for Theoretical Physics,  Universiteit  Leiden,
P.O. Box 9506, 2300 RA Leiden, The Netherlands
}
\date{\today}

\begin{abstract}
We show that finite size superconductors have a spectrum of states at extremely low energy, i.e. inside the superconducting gap. The presence of this {\it thin spectrum} is a generic feature and related to the fact that in a superconductor the global phase rotation symmetry is spontaneously broken.  For a strong coupling superconductor we find the spectrum by exactly solving the Lieb-Mattis type Hamiltonian onto which the problem maps. One of the physical consequences of the presence of thin states is that they cause quantum decoherence in superconducting qubits of finite extent.
\end{abstract}

\maketitle

{\it Introduction}
A most striking theorem in gauge theory was proved three decades ago by Elitzur, stating that local (gauge) symmetries cannot be broken spontaneously~\cite{Elitzur75}. The consequence of this no-go theorem is that the quantum mechanical expectation value of any local, non gauge-invariant operator must vanish. Naively applying the theorem to superconductors leads to the disconcerting conclusion that the pair amplitude $\langle \psi^\dagger({\bf r}) \psi^\dagger({\bf r}) \rangle$ --a manifestly non gauge-invariant quantity-- vanishes for any position $\bf r$. Taken at face value it appears that the picture of a superconductor as a state characterized by an orderparameter with a well-defined phase and amplitude breaks down.  Recently this observation motivated Hansson and coworkers to explore the scenario that superconductors are instead topologically ordered~\cite{Hansson04}.

But there is also a much more pedestrian resolution to the paradox that is posed by Elitzur's theorem. It is well known that the theorem does not forbid spontaneous breaking of the {\it global} part of the symmetry.  We exploit this caveat to construct explicitly the symmetry broken groundstate for a strong coupling superconductor. This state is characterized by a well-defined {\it global} complex-valued orderparameter and is at the same time manifestly invariant under {\it local} gauge transformations. The appearance of the Meissner effect indicates that our symmetry broken state is indeed a superconductor. In spite of the broken phase symmetry being a continuous one, there are no low-energy Goldstone modes associated with it. A well-defined energy gap develops because the Goldstone modes conspire with the electro-magnetic gauge field to create massive excitations --the well-known Anderson-Higgs mechanism for mass generation.

We will prove, however, that not all states become gapped. A spectrum of states at extremely low energy, inside the superconducting gap, remains. These in-gap states are associated with the breaking of a global symmetry in quantum systems and known in the context of antiferromagnets as the {\it thin spectrum}. In macroscopic superconductor the thin spectrum is very dense and difficult to observe, but in finite size systems the spectrum displays small, a priori observable, energy gaps. Demonstrating their presence in any superconductor is the main result of this Letter.  This has important physical consequences, one of which is a limit to the coherence of Cooper-pair box qubits.

Below we start by constructing the symmetry broken state of a strong coupling superconductor, in which electrons form local pairs. The advantage of the local pairing Hamiltonian is that it provides a manifestly gauge invariant description of a superconductor. It also allows us to solve the resulting symmetry breaking Hamiltonian exactly by directly mapping it onto a modified Lieb-Mattis model. The fact that we consider this particular strong coupling model for a superconductor does not the affect the generality of our results on spontaneous symmetry breaking (SSB) and the associated thin spectrum. As there is no phase transition between strong and weak coupling superconductivity, these systems are identical from the point of view of symmetry. Or, in more specific terms: the thin spectrum emerges from a {\it global} symmetry breaking and is therefore independent of the microscopic form and range of {\it local} interactions between electrons.

{\it Local Pairing Superconductor}
By means of the negative $U$ Hubbard Hamiltonian in presence of an electro-magnetic vector potential we investigate how SSB occurs in superconductors. This Hubbard Hamiltonian has been studied extensively~\cite{Micnas90} and it is well known to have a superconducting groundstate. It reads
\begin{eqnarray}
H &=& t \sum_{j,\delta,\sigma} \left( e^{i  \psi^{\delta}_j}
c^{\dagger}_{j+\delta,\sigma} c^{\phantom \dagger}_{j,\sigma} +  h.c.  \right) \nonumber \\
&& -\left| U \right| \sum_j n_{j,\uparrow} n_{j,\downarrow}
- \mu \sum_{j,\sigma} n_{j,\sigma} +H_{EM},
\label{Hzen}
\end{eqnarray}
where $c^{\dagger}_j$ creates an electron on site $j$, $\delta$ connects neighboring sites on a cubic lattice and $n_j$ counts the number of electrons. The amplitude of the hopping integral is $t$, $-|U|$ the strength of the local attraction between electrons and $\mu$ the chemical potential, which in general breaks particle-hole symmetry. The coupling of the charges to the vector potential comes about via the Peierls construction: while hopping from $j$ to $j+\delta$ an electron picks up a phase  $\psi^{\delta}_j$ that is proportional to the electromagnetic vector potential integrated along the bond:  $\psi^{\delta}_j=\frac{e}{\hbar c} \int_j^{j+\delta} A^{\delta}(l) dl$. A gauge transformation amounts to simultaneously sending ${\bf A} \rightarrow {\bf A} + \nabla \Lambda$ (equivalently: $ \psi^{\delta}_j \rightarrow \psi^{\delta}_j + \Lambda_{j+\delta} -\Lambda_j$) and  $c_j \rightarrow  e^{i  \Lambda_j} c_j$. The Hamiltonian is manifestly gauge invariant. We explicitly include the free electro-magnetic field in the Hamiltonian.  In the next stage of the calculation this term is needed to generate a mass for the Goldstone modes. 

In the strong coupling limit, where $U \gg t$, the electrons form strongly bound pairs and for the low-energy dynamics of the system we can restrict ourselves to the lowest Hubbard sector. In that sector sites are either empty, or doubly occupied. Sites with single electrons are only virtually allowed which gives rise to pair-pair interactions.  The effective pair dynamics can be computed from a second order perturbation expansion and is given in terms of pseudospin $1/2$ operators as
\begin{eqnarray}
H_S &=& J \sum_{j,\delta} {\frac{1}{2}}\left(
e^{-2i \psi_{j}^{\delta}} S^+_j  S^-_{j+\delta} + e^{2i \psi_{j}^{\delta}} S^-_j  S^+_{j+\delta} \right)  \nonumber \\
&& + J \sum_{j,\delta}  S^z_j S^z_{j+\delta} - h \sum_j  S_j^z +H_{EM},
\label{eq:H_S}
\end{eqnarray}
where $J=2 t^2/|U|$, and $S^+_j \equiv c^{\dagger}_{j,\uparrow} c^{\dagger}_{j,\downarrow}$. The overall pair density is given by the $z$ component of the total pseudospin $S^z_{tot}$, which can be varied by changing the parameter $h \equiv |U|-2\mu$. Away from half filling, when $h \neq 0$, the global SU(2) symmetry of the Hamiltonian is broken down to a U(1) symmetry that describes the collective rotation of the pseudospins around the $z$-axis.

Note that the prefactor of both the first and second term in the Hamiltonian above is $J$. This symmetry is removed when long-range Coulomb interactions between electron pairs are included into the model, having the form  $\sum_{j,{\bf r}} V({\bf r})  S^z_j S^z_{j+{\bf r}}$. For the discussion that follows, however, this is an inconsequential detail as such interactions preserve the global U(1) symmetry.

The Hamiltonian above can be simplified by introducing another set of pseudospins $\sigma$ that absorb the vector potential $\psi_j^{\delta}$. In this procedure, which is also used in weak coupling superconductivity, one defines  the new pseudospins as $\sigma^z_j = S^z_j$ and $\sigma_j^+ = e^{-2i \sum_{j'=0}^j \psi_{j'}^{\delta}} S^+_j$. The sum over $j^\prime$ that occurs in the definition of $\sigma^+$ is a sum over the vector potentials along an arbitrary path that connects site $j$ to an arbitrarily chosen origin at $j=0$. It is easy to show that due to flux quantization this sum does not depend on the specific path. With this mapping the effective Hamiltonian for the local pairs reduces to an antiferromagnetic Heisenberg model in a magnetic field
\begin{eqnarray}
H_{\sigma} &=& J \sum_{j,\delta} \vec{\sigma}_j^{\phantom z} \cdot \vec{\sigma}_{j+\delta}^{\phantom z}
- h\sum_j \sigma_j^z +H_{EM}.
\label{eq:H_sigma}
\end{eqnarray}
Notice that the $\sigma$ pseudospin operators by themselves are not gauge invariant. They implicitly contain the vector potential and under a general local gauge transformation all $\sigma$-pseudospins pick up the same global phase factor $e^{2i \frac{e}{\hbar c} f(0)}$. The Hamiltonian as a whole is, of course, still fully gauge invariant.
Following the study of  spontaneous symmetry breaking in magnets and crystals~\cite{vanWezel06} we split up the Hamiltonian above into two parts, a finite momentum ($\bf k$) sector and a collective, zero momentum sector. In the antiferromagnet the finite $\bf k$ sector contains the spinwave Goldstone modes which in a superconductor become massive. These modes are disjunct from the zero momentum, collective sector which contains the thin spectrum that is needed for SSB.

\begin{figure}
\includegraphics[width=0.8\columnwidth]{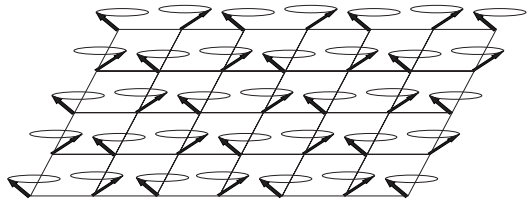}
\caption{A schematic representation of the groundstate of the local pairing superconductor on a square lattice. The arrows are semiclassical representations of the pseudospins ${\bf \sigma}$.}
\label{cantedAFM}
\end{figure}

{\it Pseudospinwaves} First we consider the finite momentum (pseudo-)spinwave excitations of the Hamiltonian above, for which a semiclassical treatment suffices. Here we only give a brief account as the resulting physics is well known. The main point is that our Hamiltonian properly incorporates the Andersson-Higgs mechanism, generating a gap for the pseudospinwave modes~\cite{Anderson63,Higgs64:1,Higgs64:2}.

The  groundstate of $H_{\sigma}$ is determined by the competition between the first and second term in the Hamiltonian. The field $h$ tends to align the spins along the $z$ axis, whereas the  interaction term in the Hamiltonian favors antiferromagnetic alignment of the pseudospins. The net result is, in terms of semiclassical pseudospins, a canted antiferromagnet in which all spins have equal projections on the $z$-axis, but are antiferromagnetically ordered in the $xy$ plane (Fig.~\ref{cantedAFM}). A low energy excitation corresponds to a long wavelength rotation of pseudospins around the $z$-axis, which amounts to a slow variation in the angle  $\phi_j^\sigma$ of each $\sigma$-pseudospin $j$ in the $xy$ plane. According to Hamiltonian~(\ref{eq:H_S}) the corresponding spinwave energy  is proportional to $\sum_{j, \delta} J \cos (2 \psi_j^{\delta} + \phi_j^S - \phi_{j+\delta}^S) $, which in the continuum limit and for small rotations reduces to an excitation energy of  $ \frac{J}{2} ({\bf A}-\nabla \phi)^2 $. The important point is that the pseudospinwave excitations in the superconductor are manifestly coupled to the vector potential. This is made explicit by introducing a transformed vector potential $\bf \tilde{A}= {\bf A}-\nabla \phi$ that absorbs the phase rotations. Gauge invariance requires the electro-magnetic part of the Hamiltonian in terms of  $\bf \tilde{A}$ and ${\bf A}$ to be identical, so that the total Hamiltonian that governs the  elementary excitations becomes $ \frac{J}{2}{\bf \tilde{A}}^2+H_{EM}({\bf \tilde{A}})$. For any finite $J$ the elementary excitations of this Hamiltonian (which are combined excitations of the electromagnetic field and the pseudospinwaves) are massive. Via this Andersson-Higgs mechanism a gap is generated for all finite-momentum pseudospinwave excitations in our superconductor. The Meissner effect is a direct physical consequence of this mass.

{\it Thin Spectrum and SSB}
The semiclassical approach above considers variations in the relative angles $\phi_j$ and $\phi_{j+\delta}$ between neighboring pseudospins.  The absolute angle of the pseudospins, however, is arbitrary. Consequently no unique classical groundstate exists: if all spins are rotated simultaneously around the $z$-axis by the same angle a different classical state results, whereas the classical groundstate energy is invariant under such a rotation. In a proper quantummechanical treatment, however, the groundstate is both rotationally invariant and unique. In the following we will show that this rotational invariance of the quantum system can be spontaneously broken due to the presence of a thin spectrum.

In exact analogy with the description of spontaneous symmetry breaking in for example crystals and antiferromagnets~\cite{vanWezel06}, the collective part of the antiferromagnetic pseudospin Hamiltonian~(\ref{eq:H_sigma}) is given by its $k=0$ and $k=\pi$ part, leading to
\begin{eqnarray}
H_{coll} = \frac{4 J}{N} \vec{\sigma}_A^{\phantom z} \cdot \vec{\sigma}_B^{\phantom z} - h \sigma_{tot}^z ,
\label{eq:Hcoll}
\end{eqnarray}
where we have defined two sublattices $A$ and $B$ between which the spins have antiparallel projections on the $xy$-plane. The Hamiltonian above is the Lieb-Mattis Hamiltonian in presence of a uniform field $h$ and can easily be diagonalized by introducing the total spin $\vec{\sigma}_{tot}= \vec{\sigma}_A +\vec{\sigma}_B$. The groundstate is non-degenerate and characterized by the total spin quantum number $\sigma_{tot}$ and its $z$-projection $\sigma_{tot}^z$. As in the antiferromagnet, the groundstate does not break the rotational invariance of its governing Hamiltonian. It is easy to see that also a set of excited states exists that have an energy $J/N$ higher than the groundstate. These states form the thin spectrum of the symmetry unbroken Hamiltonian.

To break the symmetry an external field needs to be added to the Hamiltonian. Such a field is generated by weakly coupling our system to a second, external superconductor. This leads to an effective Hamiltonian for our system of the form $H_{coll}^{SB}=H_{coll}- B \left(\sigma_A^x -\sigma_B^x \right)$, where $B$ is the effective symmetry breaking field due to the external superconductor. It should be noted that in spite of the fact that the coupling between the our superconductors and the external one is a gauge invariant quantity, the effective symmetry breaking field that appears in $H_{coll}^{SB}$ is not. The great advantage that this Hamiltonian can be solved exactly~\cite{vanWezel06,Kaiser89,Kaplan90} counterweights the small complication due to the introduction of an implicit gauge fix. As under a gauge transformation the direction of the symmetry breaking field rotates around the $z$-axis, different states that are related to each other by such a rotation make up a gauge volume. Although they appear to be distinct all these states in fact correspond to the same physical state. We will have to check that the thin spectrum of $H_{coll}^{SB}$ is made up out of distinct physical states, that are not in the gauge volume.

\begin{figure}[t]
\includegraphics[width=\columnwidth]{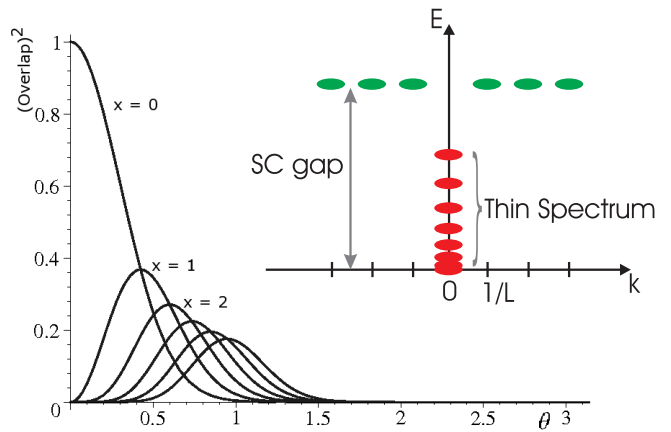}
\caption{The overlap between the thin spectrum state $\left| n \right>$ and the rotated groundstate $\hat{R}\left( \theta\right) \left| 0 \right>$, as a  function of the angle of rotation $\theta$, for different values of $n$. To make this graph we used the values  $J=10$, $B=h=1$ and $N=100$. For higher values of $N$ the graph for each $n$  will be scaled horizontally, but the height of the top remains  unaffected. The plot is symmetric under mirroring in the $\theta=0$ axis.
The inset is a schematic representation of the dispersion relation of the low-energy, low-momentum states of a finite superconductor: the states at finite $k$ are gapped, while the thin spectrum states at $k=0$ are within the gap.}
\label{overlap}
\end{figure}

The symmetry broken eigenstates of the local pairing superconductor are linear combinations of total spin states. The coefficients of these wavefunctions are given by Hermite polynomials, similar to the case of a regular antiferromagnet~\cite{vanWezel06}. The groundstate that is constructed in this manner has a specific absolute total phase determined by the symmetry breaking field. Indeed it corresponds directly to the classically realized superconducting groundstate that we considered before. The states corresponding to the higher order Hermite polynomials are extremely low in energy, and in fact collapse onto the groundstate in the thermodynamic limit~\cite{vanWezel05,vanWezel06,Zaanen96,Kaplan90,Kaiser89}. They are thus easily recognized as being the thin spectrum of the symmetry broken local pairing superconductor.

Let us now address the complication that together with the symmetry breaking field $B$ we introduced an implicit gauge fix. One can wonder whether in this situation the thin spectrum states are still as physical as they are in the symmetry unbroken Hamiltonian~(\ref{eq:Hcoll}) -- the gauge fixing may have downgraded the thin spectrum states into the gauge volume of the superconducting groundstate.
We can easily check that nothing of this kind happens. With the explicit expressions for the eigenfunctions of $H_{coll}^{SB}$ we can evaluate the overlap between the thin spectrum state $\left| n \right>$ and the superconducting groundstate
$\left| 0 \right>$ that is rotated over an angle $\theta$. If $\left| n \right>$ would be in the gauge volume of $\left| 0 \right>$, the overlap should become unity for a certain rotation angle $\theta$. As shown in figure~(\ref{overlap}) this overlap is equal to unity if and only if both $x$ and $\theta$ are zero. This demonstrates that an excited thin spectrum state is not merely a global rotation of the groundstate. Therefore it is not in the groundstate's gauge volume.

{\it BCS superconductors}
In weak coupling superconductors the physical picture for symmetry breaking stays the same. An exactly equivalent qualitative description of the thin spectrum of a superconductor can be given within the BCS model~\cite{TBP}. However, it is more challenging to obtain analytical results and closed expressions in this case.
The physical picture comes to the fore most clearly if one follows Anderson by writing the standard BCS Hamiltonian in momentum space in terms of pseudospins~\cite{Anderson58}. In the superconductor the pseudospins form a domainwall structure around $k=k_F$, which separates two ferromagnetically ordered regions with opposite $z$-axis projections. Spontaneous symmetry breaking then orients the $xy$ projection of spins in the domain wall along a specific direction in the plane. Again a thin spectrum is associated with this breaking of rotation symmetry.

{\it Decoherence}
It is known that the presence of a thin spectrum in mesoscopic spin qubits leads to quantum decoherence~\cite{vanWezel05,vanWezel06}. The thin spectrum that we have now identified in superconductors is therefore expected to lead to a finite coherence time of qubits based on superconducting material. One such type of qubit that is experimentally realized is the Cooper-pair box qubit~\cite{Nakamura99,Makhlin01,Nakamura02}. In these Cooper-pair boxes a superconducting island can be brought into a superposition of having $\bar{N}$ and $\bar{N}+1$ Cooper-pairs present. Superpositions of this type can reach coherence times of up to 500 ns~\cite{Vion02,Siddiqi06}.

In the formalism that is outlined above it is easy to consider the superposition state of the superconductor which corresponds to the experimental one~\cite{TBP}. After computing the exact time evolution of such a qubit, and tracing over the unobservable thin spectrum states, we find that the coherence of the Cooper-pair box qubit decays over time. The resulting maximum coherence time we find to be $t_{spon} =  \pi \hbar  \bar{N}/k_B T$. The calculation is analogous to the one for decoherence in antiferromagnets~\cite{vanWezel06}, but for the superconductor $\bar{N}$ signifies the average number of Cooper pairs on the superconducting island. 
Just as in the cases of crystals and antiferromagnets, the details of the model (e.g. $J$ or $h$) do not enter into the expression for the maximum coherence time, which thus appears as a universal timescale~\cite{vanWezel05,vanWezel06}.
Using from experiment the values $\bar{N} \simeq 10^6$ and $T \simeq 40$ mK~\cite{Nakamura02}, we find an upper limit to the coherence time of $\simeq 0.5$ ms for the experimentally realized Cooper-pair boxes.
Clearly this timescale that is set by the presence of the thin spectrum states is much larger than the timescale that is the current limit to coherence of the Cooper-pair boxes due to other environmental factors. However, it is well possible that the limit set by the thin states will come within the experimental reach in the near future, either because the isolation from external sources of decoherence will be developed further, or because the size of the Cooper-pair box itself is reduced even more.

{\it Conclusions}
We have shown that the superconducting state is the result of spontaneous symmetry breaking. The symmetry that is broken is a global U(1) phase symmetry, which does not contradict Elitzur's theorem. In fact the resulting symmetry broken, superconducting groundstate is still fully invariant under local gauge transformations. Associated with the spontaneous symmetry breaking in superconductors is a thin spectrum of states at extremely low energies within the superconducting gap. The presence of these low energy, global excitations leads decoherence of qubits based on superconducting materials. The maximum time that such a qubit can maintain its coherence is given by $t_{spon}= \pi \hbar {\bar N} / k_B T$, where $\bar N$ is the average number of Cooper-pairs involved. This timescale is universal in the sense that it does not depend on the underlying model parameters. With the great technological advances in manufacturing superconducting qubits it can be expected that this fundamental timescale is observed in the near future.

{\it Acknowledgments}
We thank Jan Zaanen for numerous stimulating discussions and gratefully acknowledge support from the Dutch Science Foundation FOM.


\begin{thebibliography}{10}

\bibitem{Elitzur75}
S. Elitzur, Phys. Rev. D {\bf 12},  3978  (1975).

\bibitem{Hansson04}
T. Hansson, V. Oganesyan, and S. Sondhi, Ann. Phys. {\bf 313},  497  (2004).

\bibitem{Micnas90}
R. Micnas, J. Ranninger, and S. Robaszkiewicz, Rev. Mod. Phys. {\bf 62},  113
  (1990).

\bibitem{Anderson63}
P. Anderson, Phys. Rev. {\bf 130},  439  (1963).

\bibitem{Higgs64:1}
P. Higgs, Phys. Lett. {\bf 12},  132  (1964).

\bibitem{Higgs64:2}
P. Higgs, Phys. Rev. Lett. {\bf 13},  508  (1964).

\bibitem{vanWezel06}
J. van Wezel, J. Zaanen, and J. van~den Brink, Phys. Rev. B {\bf 74},  094430
  (2006).

\bibitem{Kaiser89}
C. Kaiser and I. Peschel, J. Phys. A {\bf 22},  4257  (1989).

\bibitem{Kaplan90}
T. Kaplan, W. von~der Linden, and P. Horsch, Phys. Rev. B {\bf 42},  4663
  (1990).

\bibitem{vanWezel05}
J. van Wezel, J. van~den Brink, and J. Zaanen, Phys. Rev. Lett. {\bf 94},
  230401  (2005).

\bibitem{Zaanen96}
J. Zaanen, {\em The Classical Condensates: from crystals to Fermi-liquids}
  (Leiden University, Leiden, 1996).

\bibitem{TBP}
J. van Wezel and J. van~den Brink, manuscript in preperation (2007).

\bibitem{Anderson58}
P. Anderson, Phys. Rev. {\bf 112},  1900  (1958).

\bibitem{Nakamura99}
Y. Nakamura, Y.~A. Pashkin, and J.~S. Tsai, Nature {\bf 398},  786  (1999).

\bibitem{Makhlin01}
Y. Makhlin, G. Sch\"on, and A. Shnirman, Rev. Mod. Phys. {\bf 73},  357
  (2001).

\bibitem{Nakamura02}
Y. Nakamura {\it et al.} 
Phys. Rev. Lett. {\bf  88},  047901  (1999).

\bibitem{Vion02}
D. Vion, A. Aassime, A. Cottet, P. Joyez, H. Pothier, C. Urbina, D. Esteve, and
  M.~H. Devoret, Science {\bf 296},  886  (2002).

\bibitem{Siddiqi06}
I. Siddiqi {\it et al.}
Phys. Rev. B {\bf 73},  054510  (2006).

\end{thebibliography}

\end{document}